\documentstyle[psfig,prd,aps,psfig]{revtex}

\newcommand{\Sslash}[1]{ \parbox[b]{0.6em}{$#1$} \hspace{-0.55em}
                         \parbox[b]{0.55em}{ \raisebox{-0.2ex}{$/$}}}          \newcommand{\Qslash}[1]{ \parbox[b]{1em}{$#1$} \hspace{-1.0em}
                         \parbox[b]{0.55em}{ \raisebox{0.2ex}{$/$}}}           \newcommand{\psla}{\Sslash{p}}
\newcommand{\qsla}{\Qslash{q}}

\def\be{\begin{equation}}
\def\ee{\end{equation}}
\def\bea{\begin{eqnarray}}
\def\eea{\end{eqnarray}}

\begin{document}

\title{A polarized version of the CCFM equation for gluons}
\author{Martin Maul}

\address{
Department of Theoretical Physics, 
Lund University,
S\"olvegatan 14A,
SE - 223 62 Lund, Sweden}

\newpage
\date{\today, LU-TP-01-35}
\maketitle

\begin{abstract}
A derivation for a polarized CCFM evolution
equation which is suitable to describe the scaling behavior of the
the unintegrated polarized gluon density is given. We discuss
the properties of this polarized CCFM equation and compare it
to the standard CCFM equation in the unpolarized case. 
\end{abstract}



\section{Introduction}
An understanding of the dynamics of the gluon inside the nucleon
is one of the key issues in Quantum Chromodynamics. Especially
the behavior of the unpolarized gluon distribution function 
at small momentum fractions 
has been intensively discussed over the years.
One suggested mechanism for the gluon dynamics is given by the
CCFM equation 
\cite{Ciafaloni:1988ur,Catani:1990yc,Catani:1990sg,Marchesini:1995wr},
where the dominance of factorized diagrams in 
strongly ordered soft gluon emission is used.
Recently, it has
been shown by using a Monte Carlo implementation that the CCFM equation
gives a  good description of a variety of processes in deep 
inelastic scattering
at HERA ranging from small
to large momentum fraction $x$ 
such as forward jet cross sections, high $p_\perp$ particle
spectra, charm and bottom production \cite{Jung:2001hk}. 
\newline
\newline
A completely different field where the gluon dynamics at small $x$ should
give rise to many interesting features is polarized deep inelastic
scattering. Here one is 
interested in the way the spin is distributed among quarks and gluons
in the nucleon.
Small-$x$ effects in polarized deep inelastic scattering raise considerable
interest \cite{Radel:2001iq,Badelek:2001mz} due to the possibility that one may
access the region of $x<10^{-3}$ in future projects such at THERA
\cite{Badelek:2001mz}.
Small-x contributions to polarized structure functions have 
been regarded 
in terms of limits of the standard DGLAP evolution
equation 
\cite{Ball:1995ye,Gehrmann:1996ut}. Beyond this  the double
logarithmic contributions 
\cite{Bartels:1995iu,Bartels:1996wc,Bartels:2001zs,Kotlorz:2001iu} 
and their resummation 
\cite{Blumlein:1995jp,Blumlein:1996dd,Blumlein:1996hb}
have been investigated.
A complete formalism incorporating DGLAP and $\ln^2(1/x)$ resummation
has been given in\cite{Badelek:1998nz,Kwiecinski:1999sk}.
\newline
\newline
In this contribution I will follow the principles discussed in 
\cite{Catani:1990sg} to derive a polarized version of the CCFM equation
for gluons. Here the unintegrated polarized gluon distribution enters,
which is discussed in Sec.~\ref{sec2}. 
After a presentation of the kinematical
variables in Sec.~\ref{sec3} a discussion of the principles of the soft
gluon factorization applicable in the polarized case is given 
in Sec.~\ref{sec4}.
The derivation of the polarized CCFM equation for gluons can 
be found in Sec.~\ref{sec5}, which is followed by a comparison to 
the unpolarized CCFM equation in Sec.~\ref{sec6}. In order to 
be able to discuss finally the properties of this equation we consider a 
single iterative step in the solution of this equation which can
be physically identified with dressed single gluon emission in
Sec.~\ref{sec7} and \ref{sec8}.

\section{The physical interpretation of the polarized 
unintegrated gluon density}
\label{sec2}
In traditional spin-physics polarized gluons have been mostly considered
as on-shell partons. 
As an on shell parton the gluon carries only 
two polarization states which lie parallel or antiparallel to the spin
of the nucleon where the gluon is sitting.
\newline 
\newline
In CCFM one deals, however, with off-shell gluons which can have
three polarization states. We can decompose them again in terms of the 
spin states of the underlying nucleon. 
The kinematic situation in longitudinally polarized deep inelastic scattering
(DIS) is shown in Fig.~\ref{spins}. The total hadronic cross section
$\sigma_{h\; h_p h_e}$
depends in the high energy limit on the helicity state of the electron
$h_e$ and the helicity state of the proton $h_p$. In terms of the $k_\perp$
factorization it is a convolution of the unintegrated gluon density 
$g_{h_p h_g}$ and
the 'partonic' off-shell cross section $\sigma_{p\; h_g h_e}$, 
where the spin state $h_g$
of the gluon entering the box-graph comes in. 
Then we get for the cross sections
of the two experiments where the spin-vectors
of proton and electron one time lie
parallel and one time  anti-parallel to each other:
\begin{figure}
\centerline{\psfig{figure=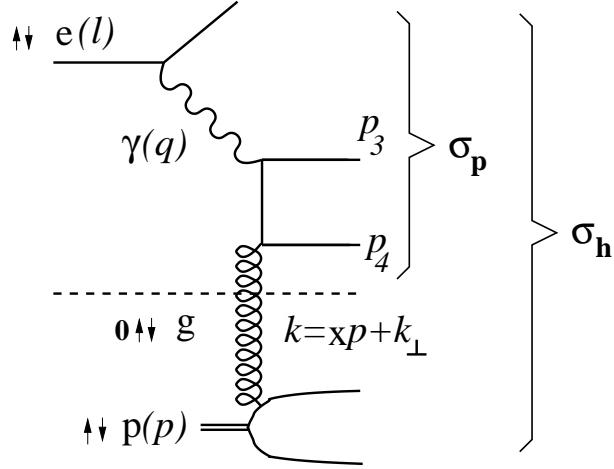,width=8cm}}   
\caption{$k_\perp$ factorization in case of polarized deep 
inelastic scattering (DIS) in terms of the 
unintegrated gluon density. The total hadronic cross section 
$\sigma_h$ factorizes into the unintegrated gluon density and 
a partonic cross section $\sigma_p$.}
\label{spins}
\end{figure}
\begin{eqnarray}
\sigma_{h\; \uparrow \uparrow} &=& 
 g_{\uparrow \uparrow}   \otimes   \sigma_{p\; \uparrow   \uparrow}
+g_{\uparrow 0}          \otimes   \sigma_{p\; 0          \uparrow}
+g_{\uparrow \downarrow} \otimes   \sigma_{p\; \downarrow \uparrow}
\nonumber \\
\sigma_{h\; \downarrow \uparrow} &=& 
 g_{\downarrow \uparrow}   \otimes   \sigma_{p\; \uparrow   \uparrow}
+g_{\downarrow 0}          \otimes   \sigma_{p\; 0          \uparrow}
+g_{\downarrow \downarrow} \otimes   \sigma_{p\; \downarrow \uparrow}\;.
\end{eqnarray}
As only relative orientations matter we can identify:
\begin{equation}
\uparrow \downarrow = \downarrow \uparrow = \times       , \qquad
\uparrow \uparrow   = \downarrow \downarrow = \parallel  , \qquad
0\uparrow = 0 \downarrow \;.
\end{equation}
So we obtain for the difference of the two hadronic cross sections:
\begin{eqnarray}
\Delta \sigma_h &=& \sigma_{h\; \times} - \sigma_{h\;\parallel}
=  \sigma_{h\; \downarrow \uparrow} - \sigma_{h\; \uparrow \uparrow}
\nonumber \\ 
&=& (g_\parallel - g_\times)\otimes 
( \sigma_{p\; \times} - \sigma_{p\;\parallel})
= \Delta g \otimes \Delta \sigma_p\;.
\end{eqnarray}
One should note that the polarization state 0 does not enter into the 
polarized cross section $\Delta \sigma_h$. Therefore, also in the unintegrated
case, the polarized gluon distribution can be defined as the difference of the
probability to find a gluon inside the proton with spin state aligned parallel
to the proton spin minus the probability to find a gluon with the corresponding
anti-parallel spin alignment. Between the
unintegrated polarized  gluon density $\Delta g(x,Q^2,k_\perp^2)$
and the integrated polarized
gluon density $\Delta g(x,Q^2)$ one has the relation:
\begin{equation}
\Delta g(x,Q^2) = \int^{Q^2} k_\perp d k_\perp  \Delta g(x,Q^2,k_\perp^2)\;.
\end{equation}
The remarkable point is that on the right hand side we have a quantity
derived from off-shell gluons while on the left hand side we have the
standard gluon parton distribution where one is thinking in terms
of on-shell gluons that have only two polarization states. It turns out indeed
that the unintegrated polarized gluon distribution fits into the decomposition
scheme of the integrated one. A systematic analysis of this topic
has been performed in Ref.~\cite{Mulders:2001sh}. The CCFM equation
I want to derive here in the polarized case is an evolution equation of
the polarized off shell unintegrated gluon density. A systematic
study of the evolution of unintegrated structure functions in terms of the 
DGLAP formalism can be found in \cite{Henneman:2001ev}.   
\section{The kinematic of the process}
\label{sec3}
\begin{figure}
\centerline{\psfig{figure=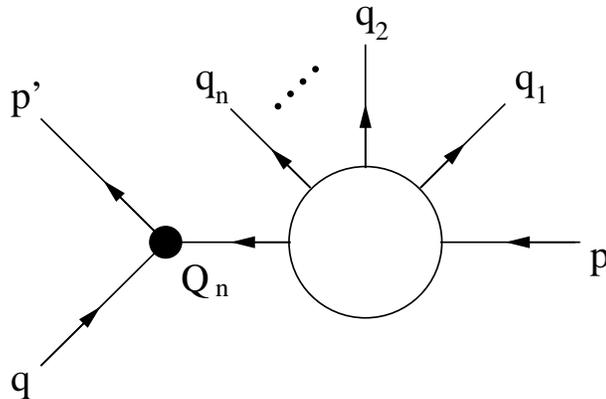,width=8cm}}   
\caption{Multi gluon emission amplitude of ordered soft gluons $q_1\dots q_n$
off the partonic gluon $p$. The hard scale is set by
$Q^2 = -q^2$, while $p'$ describes
the outgoing final gluon.}
\label{graph}
\end{figure}

As in \cite{Catani:1990sg} we study the process of parton
deep inelastic scattering, c.~f.~Fig.~\ref{graph}, where all lines,
except the one with the momentum $q$, represent gluons. 
Kinematically we have:
\begin{equation}
p + q \to p' + q_1+q_2 + \dots + q_n\;.
\end{equation}
Here $q$ acts as the hard probe of the process:
\begin{equation}
q^2 = - Q^2 < 0 , \qquad x = \frac{Q^2}{2p\cdot q}\;.
\end{equation}
The transverse momenta of the outgoing soft gluons $q_i$ are supposed to 
be smaller than the hard scale divided by $x$, i.~e.~$q_{i\perp}^2 < Q^2/x$. 
 Kinematically one introduces two light-like 
vectors:
\begin{equation}
p= E(1,0,0,1),\qquad \bar p = E(1,0,0,-1), \qquad 2p\bar p = 4E^2\;,
\end{equation}
and decomposes the other momenta through:
\begin{equation}
q = -x p+ \frac{Q^2}{x} \frac{\bar p}{2 p\cdot \bar p}, \qquad
q_i = y_i p + p\cdot q_i \frac{\bar p}{p\cdot \bar p} + q_{i\perp}\;.
\end{equation}
Then, one has the relations:
\begin{equation}
2p\cdot q_i = \frac{q_{i\perp}^2}{y_i}\;.
\label{pperprel}
\end{equation}
So that the emitted soft gluons with momenta $q_i$ are on-shell partons.
If one assumes the hard scale $Q^2$ to be large one finds furthermore:
\begin{equation}
x \approx x_n = \left(1- \sum_{i=1}^n y_i \right), \qquad  p' \sim \bar p\;.
\end{equation}
In the following a strong energy ordering is assumed:
\begin{equation}
y_1 \ll y_2 \ll \dots \ll y_n\;.
\end{equation}
Finally, the polarization vectors of the gluons involved have in the
unpolarized case the simple form:
\begin{equation}
\epsilon^{(\lambda)}_\mu(q) = g_\mu ^\lambda 
- \frac{q_\mu \eta^\lambda}{q\cdot \eta}\;,
\end{equation}
where the gauge vector $\eta$ is chosen to be $\eta=\bar p \sim p'$.
\section{The factorization of soft gluon emission in the CCFM approach}
\label{sec4}
\begin{figure}
\centerline{\psfig{figure=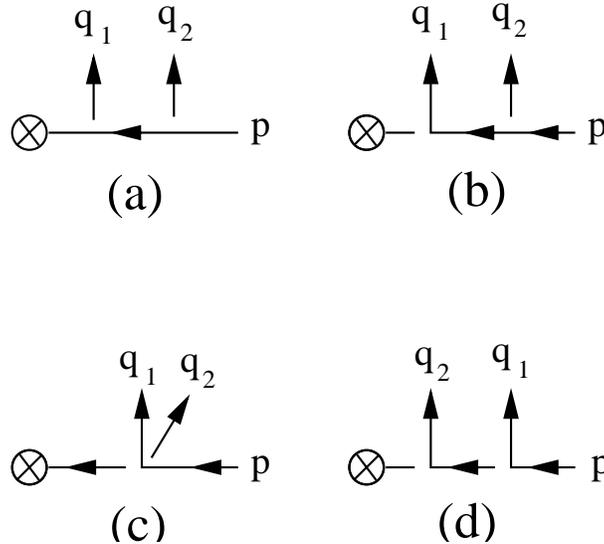,width=8cm}}   
\caption{Polarization flow for the ordered emission of two gluons
$x\ll y_1 \ll y_2 \approx 1$. Diagram (a) shows the contribution 
in the polarized case where no polarization flow is to enter the
soft gluon emission. Diagrams (b),(c) and (d) show the leading
contribution  in the unpolarized case for small $x$. Diagrams
(b) and (c) give rise to the eikonal emission while diagram (d)
is the one that accounts for the non-eikonal emission.}
\label{polflow}
\end{figure}

Lets now consider the amplitude for the soft n-gluon emission
in Fig.~\ref{graph} in terms
of $\langle acb_1 \dots b_n|M_n(p,p',q_1,\dots, q_n) \rangle$.
The basic principle of the CCFM equation is the resummation and
factorization of 
soft gluon emission. The main ingredient here is the radiation
of soft real gluons off quasi-real partons, where except for
the color matrices 
the emission vertex
is reduced to an effective scalar coupling \cite{Bassetto:1983ik}: 
\begin{eqnarray}
q(p) \to q(p-q)+g(q)\quad  &:& \quad  \frac{(\psla-\qsla)
\gamma_\mu \psla}{(p-q)^2+i\epsilon}
\to -\frac{\psla}{pq-i\epsilon}
\left(p_\mu +{\cal O}(|\vec q|)\right)
\nonumber \\
g(p) \to g(p-q)+g(q) \quad &:& \quad 
 d^{\lambda\lambda'}(p-q)
\Gamma_{\lambda'\mu \nu'}(p-q,q,p) d^{\nu\nu'}(p)
 \to 2p_\mu d^{\lambda \nu}(p)\left(1   +{\cal O}(|\vec q|)\right)\;.
\end{eqnarray}
Here $d_{\mu\mu'}(q) = - \epsilon_{\mu (\lambda)}(q) 
\epsilon_{\mu'}^{(\lambda)} (q)$ and $\Gamma$ denotes the three-gluon vertex.
The factorization of the multi-gluon emission amplitude can  be 
written in terms of a scalar current ${\bf J}_{\rm tot}^{(n-1)}$ which
consists in the unpolarized case for small $x$
of an eikonal part and a non-eikonal part. The eikonal 
part is inherited from the situation in QED where the soft photon 
emission factorizes exactly using the eikonal identity 
(c.f. Ref.~\cite{Bassetto:1983ik}):
\begin{equation}
\sum_{\rm perm} \frac{1}{a_1} \frac{1}{a_1+ a_2} \dots  
\frac{1}{a_1+ a_2+\dots a_n} = \prod_{i=1}^{n} a_i^{-1}\;.
\end{equation}
Using these principles one finds in the limit $x\to 0$ the following
iterative factorization of the soft gluon emission of the softest
gluon $q_n$ from the amplitude $M_n$:
\begin{eqnarray}
M_n &=& \frac{2(Q_n-x_n p) \cdot \epsilon^{(\lambda')}(p')}{x_n Q_n^2}
\langle acb_1 \dots b_n| {\bf h_n} (p p' q_1 \cdots q_n) \rangle
\nonumber \\
 \langle acb_1\dots b_n|{\bf h_n}(pp'q_1 \dots q_n)\rangle
&\approx& g_s \langle acb_1\dots b_{n-1}|{\bf J}_{\rm tot}^{(n-1)}(q_n) 
|{\bf h_{n-1}}(pp'q_1 \dots q_{n-1})\rangle
\nonumber \\
\nonumber \\
{\bf J}_{\rm tot}^{(n-1)}(q_n) &=& {\bf J}_{\rm eik}^{(n-1)}(q_n)
+ {\bf J}_{\rm ne}(Q_n,q_n)
\nonumber \\
 {\bf J}_{\rm eik}^{(n-1)}(q_n)& =&
- \hat T_p \frac{p^{(\lambda)}}{p\cdot q}
+ \hat T_{p'} \frac{{p'}^{(\lambda)}}{p'\cdot q}
+ \sum_{l=1}^{n-1} \hat T_l  \frac{q_l^{(\lambda) }}{q_l \cdot q}
\nonumber \\
{\bf J}_{\rm ne}(Q_n,q_n) 
&=& \frac{2(Q_{n-1}-x_{n-1}p) \cdot \epsilon^{(\lambda)}(q_n)}
{Q^2_{n-1}}\hat T_{p'}
\nonumber \\
Q_{n-1} &=&Q_{n} + q_n, \quad
x_{n-1} = x_n+y_n\;.
\label{factor}
\end{eqnarray}
Here $\hat T_q$ denotes the color charge of the gluon with the momentum $q$.
In the eikonal current a polarization component is picked.
For $x\to 1$ the factorization formula holds for the full amplitude and
no non-eikonal contributions occur:
\begin{equation}
 \langle acb_1\dots b_n|{\bf M_n}(pp'q_1 \dots q_n)\rangle
\approx g_s \langle acb_1\dots b_{n-1}|{\bf J}_{\rm eik}^{(n-1)}(q_n) 
|{\bf M_{n-1}}(pp'q_1 \dots q_{n-1})\rangle \;.
\end{equation}
In the polarized case we will see  that the non-eikonal contribution is also
absent in the case $x\to 0$.
The essential point is that for the polarized
contribution we need a spin correlation between the incoming gluon 
with momentum $p$ and the outgoing gluon with momentum $p'$. This means
that the polarization flow is in no case allowed to go 
through the soft emission.
We illustrate the situation in Fig.~\ref{polflow}. Here the ordered
emission for small $x$ is shown.
In the unpolarized case the leading diagrams are given by
(b),(c) and (d). In the cases (b) and (c) the soft gluon emission
contains no polarization flow and can therefore simply be factorized
using the eikonal emission. In diagram (d) an additional contribution
arises which gives rise to a non-eikonal contribution. 
In the polarized case the only contributing spin-configuration is
the diagram (a), where the polarization vectors
 of the incoming gluon with momentum $p$ and the outgoing gluon with
momentum $p'$ are directly correlated.
The first result is therefore that in the polarized case the non-eikonal
contribution is absent and we can use the eikonal factorization in the
same way for large and for small $x$. Now we can just perform the steps
leading to the CCFM equation which  were presented in \cite{Catani:1990sg}.
First, we single out the color amplitudes:
\begin{equation}
\langle acb_1\dots b_n| \Delta {\bf M}_n\rangle = \sum_{\pi_{n+1}}
\Delta M_{n}(p q_{l_0} \dots q_{l_n}) 
2 {\rm Tr}(\lambda^a \lambda^{b^{l_0}} \dots
\lambda^{ b^{l_n}})\;,
\end{equation}
where the sum is over permutations $l_0,\dots l_n$ with $q_0=p'$, $b_0=c$.
Neglecting
all non-leading collinear and non-planar terms,
the color algebra yields $(\sigma_0 = N_c^2-1)$: 
\begin{equation}
|\Delta {\bf M}_n|^2 = \sigma_0 \left(\frac{C_A}{2}\right)^n 
\sum_{\pi_{n+1}} |\Delta  M_n(p q_{l_0} \dots q_{l_n})|^2\;.
\end{equation}
Here $C_A=N_c$ is the number of colors.
The factorization of the soft gluon emission in terms of eikonal currents
leads to the following recurrence relation:
\begin{equation}
|\Delta  M_n(\dots q_l q_n q_{l'}\dots )|^2 \approx - g_s^2
|\Delta  M_{n-1}(\dots q_l  q_{l'}\dots )|^2 (j_l(q_n)-j_{l'}(q_n))^2\;,
\quad j_l(q_n) = \frac{q_l}{q_l \cdot q_n}\;.
\end{equation}
%
%
%
The important thing is now the initial condition because this is the place
where the polarization enters. The hard splitting kernels in the 
CCFM equation should match the DGLAP splitting kernels in the limits
$z\to 0,1$.
To see this relation in an explicit
way we will work for the last step that generates the initial conditions
with the Altarelli Parisi method itself.  
For this purpose we consider  Fig.~\ref{AltarelliParisi}:
\begin{figure}
\centerline{\psfig{figure=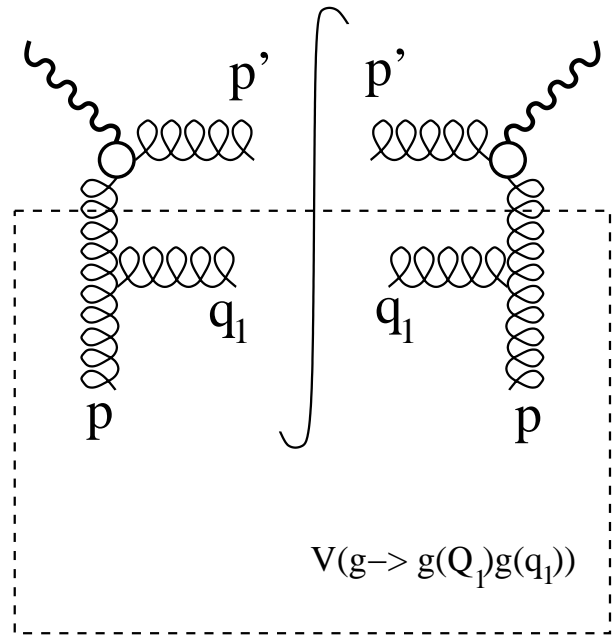,width=8cm}}  
\caption{Decomposition of the initial amplitude $M_1$ by
the Altarelly Parisi method.}
\label{AltarelliParisi}
\end{figure}
\noindent
Here the initial amplitude is decomposed in the emission of the gluon 
with momentum $q_1$ and the coupling to the hard virtuality:
\begin{equation}
|M_1(pp'q_1;\lambda\lambda')|^2 \approx \sum_{\lambda''\in \pm}
|V_{G\to GG}(p,Q_1,q_1;\lambda,\lambda'')|
|H_{G\to G}(Q_1,p';\lambda'',\lambda')|^2\;.
\end{equation}
In the Altarelli Parisi method we consider the virtuality of $Q_1^2$ to be
comparatively small, so that all gluons in the proton can
be treated as partons approximately. 
We first calculate the two gluon effective amplitude:
\begin{eqnarray}
|H_{G\to G}(Q_1,p';\lambda'',\lambda')|^2 &=& 
V_{{\rm eff}\mu\nu}(Q_1,p')
V_{{\rm eff} \mu'\nu'}(Q_1,p')
\Pi^{\mu\mu'}(x_1 p,\eta,\lambda'')
\Pi^{\nu\nu'}(p',\eta,\lambda')
\nonumber \\
&=& \frac{1}{Q_1^4}\delta_{\lambda''\lambda'}
\nonumber \\
V_{{\rm eff}\mu \nu}(Q_1,p') &=& \frac{1}{Q_1^2} \left[ -g_{\mu\nu} 
+ \frac{ p'_\mu Q_{1\;\nu}}{x_1 pp'}\right]
\nonumber \\
\Pi^{\mu\mu'}(x_1 p,\eta,\lambda'') &=&\frac{1}{2}
\left(-g^{\mu\mu'} + \frac{\eta^{\mu} p^{\mu'} +\eta^{\mu'} p^{\mu}}{\eta \cdot p}
-\lambda  \frac{i}{\eta \cdot p} \epsilon^{\mu\mu'\eta p}\right)
\nonumber \\
\Pi^{\nu\nu'}(p',\eta,\lambda') &=&\frac{1}{2}\left(-g^{\nu\nu'} 
+ \frac{\eta^{\nu} {p'}^{\nu'} +\eta^{\nu'} {p'}^{\nu}}{\eta \cdot p'}
-\lambda'  \frac{i}{\eta \cdot p'} \epsilon^{\nu\nu'\eta p'}\right)\;.
\end{eqnarray}
$H\sim \delta_{\lambda\lambda'}$ reflects the fact that the 
effective vertex cannot flip the helicity of the spin-1 gluon. To do such
a thing one would require a spin-2 particle which does not exist in 
the process. 
We should remember that in the end of the calculation we
can take the gauge vector $\eta \sim p'$. As a next step we turn to 
the calculation of the splitting amplitude $V_{G\to GG}$. Here we 
apply the Altarelli Parisi method where all  three gluons are approximately
parton like. To to this we choose the following parameterization
\cite{Altarelli:1977zs}:
\begin{eqnarray}
p &=& (P,P,{\bf 0}) \nonumber \\
Q_1 &=& (x_1 P + \frac{p_\perp^2}{2Px_1},x_1 P,{\bf p_\perp} )
\nonumber \\
q_1 &=& ((1-x_1) P + \frac{p_\perp^2}{2P(1-x_1)},(1-x_1) P,-{\bf p_\perp} )
\nonumber \\
n &=& (P,-P,{\bf 0} )\;.
\end{eqnarray}
Here $n$ acts as a gauge vector to remove unphysical decrees of freedom.
Then one obtains for the splitting amplitude leaving out the color factors
which will be provided later:
\begin{eqnarray}
|V_{G\to GG}(p,Q_1,q_1;\lambda,\lambda'')|^2 
&=&
\sum_{\lambda'''\in \pm }
|V_{G\to GG}(p,Q_1,q_1;\lambda,\lambda'',\lambda''')|^2 
\nonumber \\
&=&
2 g_s^2 p_\perp^2\left[\frac{(1-x_1+x_1^2)^2}{(1-x_1)^2x_1^2}
+ \lambda\lambda'' 
\frac{2-3x_1+2x_1^2}{(1-x_1)^2x_1}\right]
\nonumber \\
V_{G\to GG}(p,Q_1,q_1;\lambda,\lambda'',\lambda''') &=&
-g_s
\left\{
-\left[(p+q_1) \epsilon^{*b}_{Q_1,\lambda''} \right]
 \left(\epsilon^a_{p,\lambda} \cdot
  \epsilon^{*c}_{q_1,\lambda'''}\right)\right.
\nonumber \\ && \qquad \qquad 
+\left[(q_1-Q_1) \epsilon^{a}_{p,\lambda} \right]
 \left(\epsilon^{*b}_{Q_1,\lambda''} 
\cdot \epsilon^{*c}_{q_1,\lambda'''}\right)
\nonumber \\ && \qquad  \qquad
+\left.\left[(p+Q_1) \epsilon^{*c}_{q_1,\lambda'''} \right]
 \left(\epsilon^a_{p,\lambda} \cdot \epsilon^{*b}_{Q_1,\lambda''}\right)
\right\}
\nonumber \\
\left(\epsilon^a_{p,\lambda}\right)_\mu 
\left(\epsilon^{*a'}_{p,\lambda}\right)_{\mu'}
&=& \frac{1}{2} \left( - g_{\mu \mu'} 
+  \frac{ n_\mu p_{\mu'} + n_{\mu'} p_{\mu} }{ n\cdot p} +
\lambda \frac{i}{n\cdot p} \epsilon_{\mu \mu' n p} \right)\;. 
\end{eqnarray}
The next step is to put the two amplitudes together:
\begin{eqnarray}
|\Delta M_1(pp'q_1)|^2 &=& |M_1(pp'q_1;++)|^2 -  |M_1(pp'q_1;+-)|^2
= 4 g_s^2 \frac{p_\perp^2}{Q_1^4}\frac{2-3x_1 +2x_1^2}{(1-x_1)^2 x_1}\;.
\end{eqnarray}
Next we have to make use of the fact, that according to Eq.~(\ref{pperprel}):
\begin{equation}
p_\perp^2 = q_{1\perp}^2 = -2 p\cdot q_1 y_1 \approx Q_1^2 (1-x_1)\;,
\end{equation}
which in turn results in:
\begin{eqnarray}
|\Delta M_1(pp'q_1)|^2 &\approx& 4 g_s^2 
\frac{1}{Q_1^2}\frac{2-3x_1 +2x_1^2}{(1-x_1) x_1}
\nonumber \\
&\approx&   2 g_s^2 \left(\frac{2p \cdot q_1}{Q_1^2}\right)
\frac{p\cdot p'}{(p'\cdot q_1) ( p \cdot q_1)}
\frac{2-3x_1 +2x_1^2}{ x_1}
\nonumber\\ &\approx &
 g_s^2 \frac{2-3x_1 +2x_1^2}{ x_1} \left( j_p(q_1)-j_{p'}(q_1)\right)^2
\nonumber \\
&=& \left \{
\begin{array}{ccc}
 g_s^2 \frac{2}{x_1} \left( j_p(q_1)-j_{p'}(q_1)\right)^2 &  {\rm for}& (x_1\to 0) \\
 g_s^2 \left( j_p(q_1)-j_{p'}(q_1)\right)^2 &  {\rm for}&  (x_1\to 1) \\
\end{array} \right.\;.
\end{eqnarray}
Here we used that
$p'\cdot q_1 \approx (1-x_1) p p'$ and that for  
parton-like soft
gluons one can set $-2p\cdot q_1 \approx Q_1^2$.
We can interpolate the two limits using the effective expression:
\begin{eqnarray}
|\Delta M_1(pp'q_1)|^2 \equiv  g_s^2\frac{2-x_1}{x_1} \left( j_p(q_1)-j_{p'}(q_1)\right)^2\;.
\end{eqnarray}
One should note that in the 
unpolarized case this method just reproduces the results in \cite{Catani:1990sg}.
Finally,  we arrive for the amplitude at tree level to:
\begin{equation}
|\Delta M_n^{(\rm tree)}|^2 \approx \sigma_0 \left(g_s^2 C_A\right)^n
\frac{2-x_n}{x_n}
 \sum_{\pi_{n+1}}
W_n(p q_{l_0} \dots q_{l_n}), \quad
W_n(p q_{l_0} \dots q_{l_n}) = \frac{(p\cdot p')^2}{(p\cdot q_{l_0})
\dots (p\cdot q_{l_n})}\;,
\end{equation}
where $W_n$ is the multi-eikonal distribution. The result is similar to
the one obtained in the unpolarized case in Ref.~\cite{Catani:1990sg}
Eq.~(4.16) except for the factor $(2-x_n)/x_n$. We will see that
this naturally leads to the polarized DGLAP splitting kernel in the 
limit $z\to 1,0$ noting that our result is valid for small as well as
for large $z$.
\section{The derivation of the polarized CCFM equation}
\label{sec5}
The result of the previous section allows a simple and straightforward
derivation of the polarized CCFM equation. To do this we we consider
the contribution to the polarized structure function:
\begin{equation}
\sigma_0 \Delta F(Q,x) =  \Delta F_0(x) + \sum_{n=1}^\infty
\frac{1}{n!} \int \prod_{i=1}^n (d q_i) 
\Theta(Q-q_{i\perp}) |\Delta M_n|^2 \delta \left(1-\frac{x}{x_n}\right), 
\quad (dq_i)= \frac{d^2 q_i}{2\omega_i(2\pi)^3}\;.
\end{equation}
Now inserting the tree-level amplitude one obtains:
\begin{eqnarray}
\sigma_0 \Delta F^{(\rm tree)}(Q,x) &=&  
\Delta F^{(\rm tree)}_0(x) 
+ \sum_{n=1}^\infty \int \prod_{i=1}^n \frac{d\xi_i}{\xi_i} \frac{dy_i}{y_i} 
\bar \alpha_s 
\Theta(Q-q_{i\perp}) \frac{2-x_n}{x_n} \delta \left(1-\frac{x}{x_n}\right) 
\Theta^\xi_{n,\dots,2,1}
\nonumber \\
&\equiv& \Delta F^{(\rm tree)}_0(x) 
+ \sum_{n=1}^\infty \int \prod_{i=1}^n \frac{d\xi_i}{\xi_i} d z_i
\frac{\alpha_s}{2\pi} \Delta P(z_i)  
\Theta(Q-q_{i\perp}) \delta (x-z_1\dots z_n)
\Theta^\xi_{n,\dots,2,1}
\nonumber \\
\nonumber \\
\bar \alpha_s &=& C_A \alpha_s /\pi,\qquad x_n=1-y_1-\dots -y_n
\nonumber \\
x_n &=& z_1 z_2 \dots z_n , \qquad y_l = x_{l-1}(1-z_l)
\nonumber \\
\xi_i &=& \frac{p\cdot q_i}{E \omega_i}, \qquad \
\Theta^\xi_{l_k\dots l_1} = \prod_{i=1}^{k-l} 
\Theta(\xi_{l_{i+1}}-\xi_{l_{i}})\;.
\label{result1}
\end{eqnarray}
The equivalence symbol ($\equiv$) means that the two expressions become
equal in the limit $z_i\to1,0$, and one obtains as hard splitting kernel 
for the polarized CCFM equation:
\begin{equation}
\Delta P(z) =  \frac{2C_A (2-z)}{1-z} = 2C_A\left( \frac{1}{1-z}+1\right)\;.
\end{equation}
In the limit $z\to 1$ this result is identical to the pole contribution
of the corresponding polarized
DGLAP splitting function \cite{Altarelli:1977zs}:
\begin{equation}
\Delta P_{gg \; \rm pole}(z) = 
2C_A \left[ \frac{2-3z+2z^2}{(1-z)} 
+ \left(\frac{11}{12} - \frac{1}{3}\frac{T_R}{C_A}\right)\delta(1-z)\right]
\Bigg|_{\rm pole} = \frac{2C_A}{1-z}\;.
\end{equation}
Here $T_R=N_f/2$ is the number of flavors.
In the limit $z\to 0$ one obtains $4C_A$ in accordance with the corresponding
limit in the DGLAP splitting kernel. The result Eq.~(\ref{result1})
is exactly identical to the one derived in \cite{Catani:1990sg} 
Eq.~(4.29). The only difference is that we have to exchange $\Delta P(z)$
by $P(z)$ which is given by:
\begin{equation}
P(z) = 2C_A \left( \frac{1}{z} + \frac{1}{1-z}\right)\;.
\end{equation}
In the unpolarized case $P(z)$ corresponds to the pole structure
of the corresponding DGLAP splitting function for gluons $P_{gg}(z)$
\cite{Altarelli:1977zs}:
\begin{equation}
P_{gg \; \rm pole}(z) = 
2C_A \left[ \frac{z}{(1-z)_+} + \frac{1-z}{z} + z(1-z)\
+ \left(\frac{11}{12} - \frac{1}{3}\frac{T_R}{C_A}\right)\delta(1-z)\right]
\Bigg|_{\rm pole}
=
2C_A \left(\frac{1}{z} + \frac{1}{1-z}\right)\;.
\end{equation}
In polarized and unpolarized case the $1/(1-z)$ pole is the same, 
while in the polarized case the $1/z$ pole is missing.
These findings explain now simply how the 
polarized CCFM equation should be constructed:
\begin{itemize}
\item 
The hard splitting function $\Delta P(z)$ in the polarized CCFM equation
should have the form $P(z) = 2 C_A(2-z)/(1-z)$. The form is valid both
for small and for large z because the factorization in the polarized
case is valid in both limits as shown in the previous section. For 
$z\to 0,1$ this splitting function becomes identical to the polarized
DGLAP splitting function.
\item
Adding virtual corrections to the tree level results means to multiply
with the corresponding form factors as demonstrated in \cite{Catani:1990sg}.
The requirement that the polarization flow does not 
enter into the soft emission means that the non-eikonal form factor
is absent. So we have to amend only the eikonal form factor to the
tree level result. The eikonal form factor is the same as in the
unpolarized case because the polarization flow does not enter into
the soft emission. The statement means that on the average the soft
emission does not know anything of the initial spin state of the
gluon.
\end{itemize}
With these principles in mind we can write down the integral
form for the polarized CCFM equation
taking into account only  gluons:
\begin{eqnarray}
\Delta g(x,\vec k_\perp^2,Q^2) &=& 
\Delta g_0(x,\vec k_\perp^2,Q^2) 
\nonumber \\ &&
+ \int_x^1 \frac{dz }{z} 
\int_0^{2\pi} \frac{d \theta_{q_\perp'}}{2\pi}
\int_{Q_0^2}^{\infty} \frac{d{\vec {q_\perp'}}^2}{{\vec {q_\perp'}}^2}
\Theta(Q-z |{\vec {q_\perp'}}|)
\Delta^{(g)}_{\rm e}  (\bar Q^2,(z{\vec {q_\perp'}})^2)
\Delta {\cal P}_{gg}(z ,{\vec {q_\perp'}}^2, \vec k_\perp^2) 
\Delta g(x/z,{\vec {k_\perp'}}^2, {\vec {q_\perp'}}^2)
\nonumber \\
\vec k_\perp' &=& \vec k_\perp + (1-z) {\vec {q_\perp'}}\;. 
\label{mainresult}
\end{eqnarray}
Here the CCFM-kernel has the structure:
\begin{eqnarray}
\Delta{\cal P}_{\rm gg}(z, q^2,k_\perp^2) 
&=& \frac{\alpha_s(q^2(1-z)^2)}{2\pi} 
\Delta P_{gg}(z) \;.
\end{eqnarray}
The CCFM kernel in the polarized  case is $k_\perp$ independent due
to the absence of the non-eikonal form factor and the way the scale
in $\alpha_s$ is chosen. But in general it is a function of 
$k_\perp$, for example in the unpolarized case.
The eikonal form factor $\Delta_{\rm e}^{(g)}$ is taken
from Ref.~\cite{Catani:1990sg}:
\begin{eqnarray}
\Delta_{\rm e}^{(g)}(\bar q^2 , (zq)^2) &=& 
\exp\left( - \int_{(zq)^2}^{{\bar q}^2} \frac{d {q'}^2}{{q'}^2}
\int_0^{1-Q_0/{q'}} \frac{dz}{1-z} \frac{\alpha_s({q'}^2(1-z)^2)}{\pi} C_A
\right)\;,
\end{eqnarray}
and the hard splitting kernel reads: 
\begin{equation}
\Delta P_{gg}(z) = \frac{2C_A (2-z)}{1-z}\;.
\end{equation}
\section{Comparison between the polarized and unpolarized CCFM equation}
\label{sec6}
In order 
to understand the physics of the result obtained in the previous chapter
we compare it to the unpolarized CCFM equation. To make 
the comparison as instructive as possible we show the corresponding
equations here together:
In integral form the polarized and unpolarized CCFM equation reads:
\begin{eqnarray}
\Delta g( x ,\vec k_\perp^2,Q^2) &=& 
\Delta g_0( x ,\vec k_\perp^2,Q^2) 
\nonumber \\ &&
+ \int_x^1 \frac{d z  }{ z } 
\int_0^{2\pi} \frac{d \theta_{q_\perp'}}{2\pi}
\int_{Q_0^2}^{\infty} \frac{d{\vec {q_\perp'}}^2}{{\vec {q_\perp'}}^2}
\Theta(Q- z  |{\vec {q_\perp'}}|)
\Delta^{(g)}_{\rm e}  (\bar Q^2,(z{\vec {q_\perp'}})^2)
\Delta {\cal P}_{gg}( z  ,{\vec {q_\perp'}}^2, \vec k_\perp^2) 
\Delta g( x / z ,{\vec {k_\perp'}}^2, {\vec {q_\perp'}}^2)
\nonumber \\
 g( x ,\vec k_\perp^2,Q^2) &=& 
 g_0( x ,\vec k_\perp^2,Q^2)\nonumber \\ &&
 + \int_x^1 \frac{d z  }{ z } 
\int_0^{2\pi} \frac{d \theta_{q_\perp'}}{2\pi}
\int_{Q_0^2}^{\infty} \frac{d{\vec {q_\perp'}}^2}{{\vec {q_\perp'}}^2}
\Theta(Q- z  |{\vec {q_\perp'}}|)
\Delta^{(g)}_{\rm e}  (\bar Q^2,(z{\vec {q_\perp'}})^2)
 {\cal P}_{gg}( z  ,{\vec {q_\perp'}}^2, \vec k_\perp^2) 
g( x / z ,{\vec {k_\perp'}}^2, {\vec {q_\perp'}}^2)\;.
\end{eqnarray}
For the CCFM splitting kernels one has:
\begin{eqnarray}
\Delta {\cal P}_{\rm gg}(z,q^2,k_\perp^2) 
&=& \frac{\alpha_s(q^2(1-z)^2)}{2\pi} 
\Delta P_{gg }(z) 
\nonumber \\
{\cal P}_{\rm gg}(z,q^2,k_\perp^2) 
&=& \frac{\alpha_s(q^2(1-z)^2)}{2\pi} 
\Delta_{\rm ne} (z,q,k_\perp)
P_{gg \; \rm pole}(z) \;,
\end{eqnarray}
with the eikonal and non-eikonal form factors 
and hard splitting kernels given by:
\begin{eqnarray}
\Delta_{\rm e}^{(g)}(\bar q^2 , (zq)^2) &=& 
\exp\left( - \int_{(zq)^2}^{{\bar q}^2} \frac{d {q'}^2}{{q'}^2}
\int_0^{1-Q_0/{q'}} \frac{dz}{1-z} \frac{\alpha_s({q'}^2(1-z)^2)}{\pi} C_A
\right)
\nonumber \\ 
\Delta_{\rm ne} (z,q,k_\perp) &=&
\exp\left( -\frac{\alpha_s(q^2(1-z)^2)}{\pi}C_A \int_z^1
\frac{dz'}{z'} 
\int^{k^2_\perp}_{(z'q)^2} \frac{dq^2}{q^2}\right)
\nonumber \\
\nonumber \\
\Delta P_{gg }(z) &=& 2 C_A \frac{2-z}{1-z} 
\nonumber \\
P_{gg \; \rm pole}(z) &=& 2 C_A \left(\frac{1}{z} + \frac{1}{1-z}\right)\;.
\end{eqnarray}
Comparing the two equations one can say something of the
underlying physics:
\begin{itemize}
\item
In the limit of large $x$ when the non-eikonal form factor
becomes equal to unity, the CCFM equation in the unpolarized as well
as in the polarized case becomes identical to the corresponding DGLAP
evolution equation, except for the eikonal form factor. 
\item
There is no $1/z$ pole in the polarized hard splitting kernel $\Delta P(z)$
in the polarized CCFM equation in parallel
to the fact that there is no 1/z pole in the polarized DGLAP splitting
function either.
\item
In the polarized case we need to see spin correlations all along over
the way of the soft emission. For this correlation only diagrams contribute
that have no polarization flow into the soft gluons. For this reason
there are no non-eikonal contributions to the polarized CCFM equation.
This is very natural because the non-eikonal contributions 
in the unpolarized CCFM equation couple actually only to the
1/z pole, which is absent in the polarized case.
\item
In the polarized case the hard splitting function has the form $(2-z)/(1-z)$ 
which means that for small z the polarization is not enhanced by a pole
as opposite to the unpolarized case.
\end{itemize}
To resum, these findings say that soft gluon emission destroys
to a large extent
the definite polarization state of the incoming gluon and that the 
soft emission knows on the average not very much of the initial polarization
of the gluon. This result is quite remarkable because it justifies a long
termed used practice in polarized MC event generators like PEPSI 
\cite{Martin:fp}, where the
unpolarized parton showering formalism has been used to simulate the
soft gluon emission in polarized events \cite{Maul:1998si}.
\section{Single dressed gluon emission}
\label{sec7}
In principle it would be now desirable to take some input distribution
$\Delta g_0$, evolve it and compare it with data. Such a project is however
beyond the scope of this article because there are a couple of severe problems
to be solved first. Among those are:
\begin{itemize}
\item
There is no simple analytic formalism to solve the CCFM equation like
in DGLAP with the conversion into Mellin moments. Indeed a genuine
solution of the CCFM equation seems to be possible only by means 
of Monte Carlo technique.
\item
For small $z$ the hard splitting kernel for gluons
goes to a constant, as also the quark splitting functions at least 
in the DGLAP formalism do. So the influence of the quarks in the
polarized case is in principle of the same order as the gluons and
cannot be neglected.
\end{itemize}
Instead of this we want to investigate a bit more closely how the 
equation works in the polarized case. In principle a possible solution
could be obtained iteratively by:
\begin{eqnarray}
g_{n+1}( x ,\vec k_\perp^2,Q^2)
&=& g_{0}( x ,\vec k_\perp^2,Q^2) + \Delta_e(Q^2,Q_0^2)
\int_{Q_0^2}^{Q^2}\frac{d q^2}{q^2}
\int_x^1 \frac{d z  }{ z } \int_0^{2\pi}  \frac{d \theta_{q_\perp'}}{2\pi}
\frac{\Delta {\cal P}_{gg}( z  ,q^2/z^2, \vec k_\perp^2)}
{\Delta_e(q^2,Q_0^2)} 
g_n( x / z ,{\vec {k_\perp'}}^2, q^2/z^2)
\nonumber \\
g( x ,\vec k_\perp^2,Q^2)
&=&
\lim_{n\to\infty} g_{n}( x ,\vec k_\perp^2,Q^2)\;.
\end{eqnarray}
To restrict the numerical effort as much as possible we restrict ourself
to the first order of this iteration and calculate $\Delta g_1$, and 
for comparison also $g_1$. From a physical point of view this corresponds
to a single dressed gluon emission. It is a dressed emission because
the eikonal form factor includes virtual corrections to the single
soft gluon emission. In this case all necessary formulas become quite simple.
As the starting distributions we use the usual ansatz where the $k_\perp$
dependence is determined by a Gaussian, c.f.~\cite{Jung:2001hk}:
\begin{eqnarray} 
\Delta g_0( x ,\vec k_\perp^2,Q^2) &=& N  x ^\alpha(1- x )^\beta 
\frac{1}{ k_0^2}
\exp\left(-\frac{{\vec k_\perp}^2}{2k_0^2}\right)\Delta_e(Q^2,Q_0^2)
\nonumber \\
 g_0( x ,\vec k_\perp^2,Q^2) &=& N'  x ^{\alpha'}(1- x )^{\beta'} 
\frac{1}{ k_0^2}
\exp\left(-\frac{{\vec k_\perp}^2}{2k_0^2}\right)\Delta_e(Q^2,Q_0^2)\;.
\label{ansatz}
\end{eqnarray}
Using this ansatz one can perform for the single dressed gluon emission
the angle integration analytically:
\begin{eqnarray}
\Delta g_{1}( x ,\vec k_\perp^2,Q^2)
&=&  \Delta g_{0}( x ,\vec k_\perp^2,Q^2) + \Delta_e(Q^2,Q_0^2)
\int_{Q_0^2}^{Q^2}\frac{d q^2}{q^2}
\int_x^1 \frac{d z  }{ z } 
\frac{\Delta {\cal P}_{gg}( z  ,q^2/z^2, \vec k_\perp^2)}{\Delta_e(q^2,Q_0^2)}
\Delta g_0( x / z ,Q_0^2)
\nonumber\\ && \qquad \qquad \qquad \qquad \times \frac{1}{k_0^2}
\exp\left[
-\frac{\left(\frac{1- z }{ z }q \right)^2 + k_\perp^2}{2k_0^2} 
\right]
I_0 \left(\frac{(1- z )q k_\perp}{ z  k_0^2} \right)\;.
\end{eqnarray}
Here $I_0$ is the modified Bessel function of zero order as obtained from:
\begin{equation}
I_0(\alpha) = \int_0^{2\pi} \frac{d\theta}{2\pi} 
\exp(\alpha \cos \theta)\;.
\end{equation}
One should notice that this approximate solution holds for the unpolarized
case as well, so we can summarize our result for $g_1$ and $\Delta g_1$
to be:
\begin{eqnarray}
\Delta g_{1}( x ,\vec k_\perp^2,Q^2)
&=&
\Delta g_{0}( x ,\vec k_\perp^2,Q^2) + 
\Delta_e(Q^2,Q_0^2)
\int_{Q_0^2}^{Q^2}\frac{d q^2}{q^2}
\int_x^1 \frac{d z  }{ z } 
\Delta {\cal P'}_{gg}( z  ,q^2/z^2, \vec k_\perp^2)
\Delta g_0( x / z ,Q_0^2)
\nonumber \\
 g_{1}( x ,\vec k_\perp^2,Q^2) 
&=&
 g_{0}( x ,\vec k_\perp^2,Q^2) + 
\Delta_e(Q^2,Q_0^2)
\int_{Q_0^2}^{Q^2}\frac{d q^2}{q^2}
\int_x^1 \frac{d z  }{ z } 
 {\cal P'}_{gg}( z  ,q^2/z^2, \vec k_\perp^2)
g_0( x / z ,Q_0^2) \;,
\end{eqnarray}
using
\begin{eqnarray}
\Delta {\cal P}'_{gg}(z ,q^2/z^2, k_\perp^2) 
&=& \frac{\alpha_s\left(q^2\frac{(1-z)^2}{z^2}\right)}
{2\pi}
\Delta_{\rm e}(Q_0^2,q^2)
\Delta P_{\rm gg}(z)
\nonumber \\ && \qquad \qquad \qquad \times 
\frac{1}{k_0^2}\exp\left[
-\frac{\left(\frac{1- z }{ z }q \right)^2 + k_\perp^2}{2k_0^2} 
\right]
I_0 \left(\frac{(1-z)q k_\perp}{z k_0^2} \right)
\nonumber \\
 {\cal P}'_{gg}(z ,q^2/z^2, k_\perp^2) 
&=& \frac{\alpha_s\left(q^2\frac{(1-z)^2}{z^2}\right)}
{2\pi} \Delta_{\rm ne}(z,q/z,k_\perp)
\Delta_{\rm e}(Q_0^2,q^2)
 P_{\rm gg\;pole}(z)
\nonumber \\ && \qquad \qquad \qquad \times 
\frac{1}{k_0^2}
\exp\left[
-\frac{\left(\frac{1- z }{ z }q \right)^2 + k_\perp^2}{2k_0^2} 
\right]
I_0 \left(\frac{(1-z)qk_\perp}{z k_0^2} \right)\;.
\end{eqnarray}
\section{Numerical analysis of the single dressed gluon emission}
\label{sec8}

\begin{figure}
\centerline{\psfig{figure=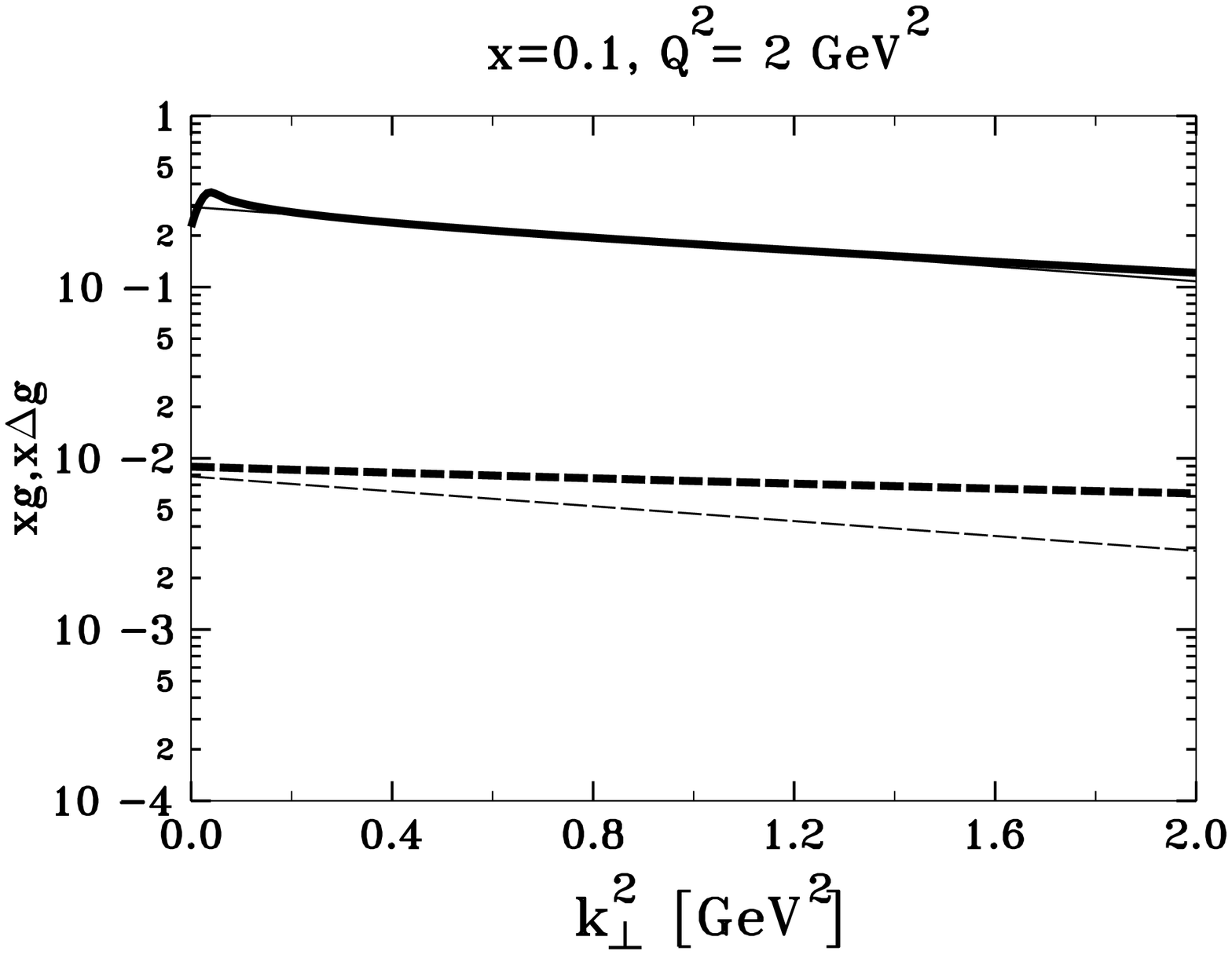,width=10cm}}
\centerline{\psfig{figure=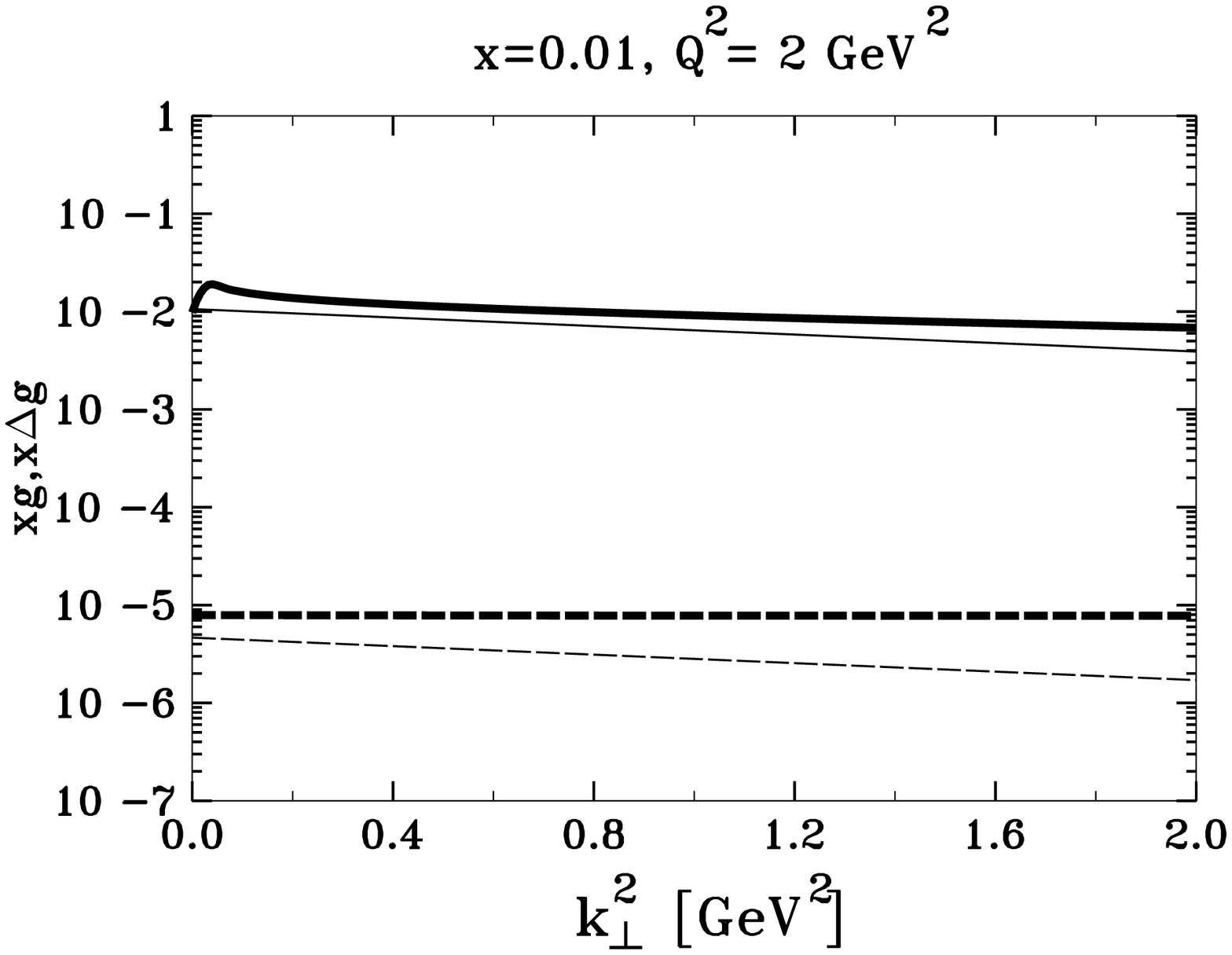,width=10cm}}  
\centerline{\psfig{figure=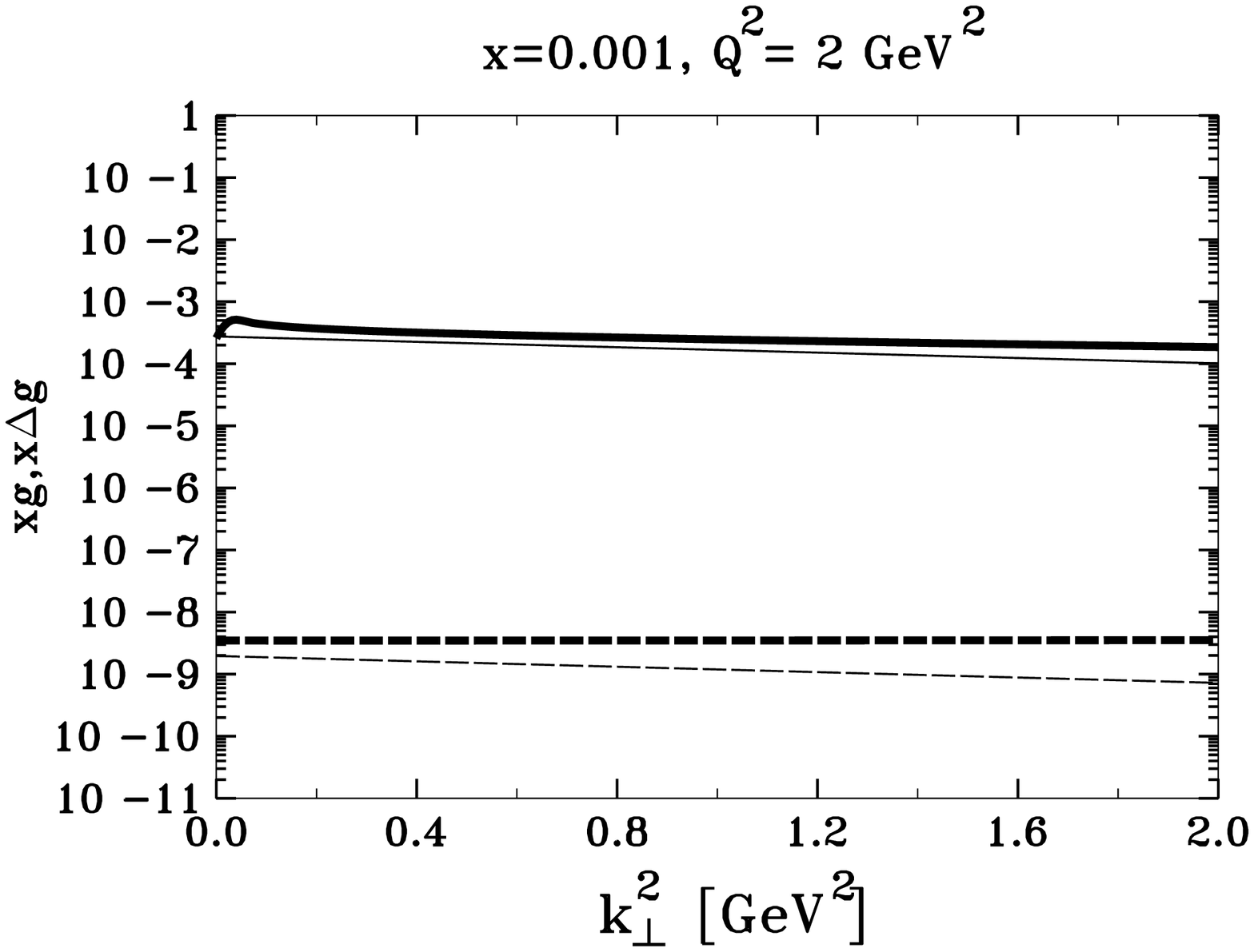,width=10cm}}  
\caption{$k_\perp$ dependence of the unintegrated gluon distribution
for $Q^2=2\;{\rm GeV}^2$:\\
bold solid line $xg_1(x,k_\perp^2)$, thin solid line $xg_0(x,k_\perp^2)$,\\
bold dashed line $x\Delta g_1(x,k_\perp^2)$, thin dashed line 
$x \Delta g_0(x,k_\perp^2)$.}
\label{kp2q22}
\end{figure}

\begin{figure}
\centerline{\psfig{figure=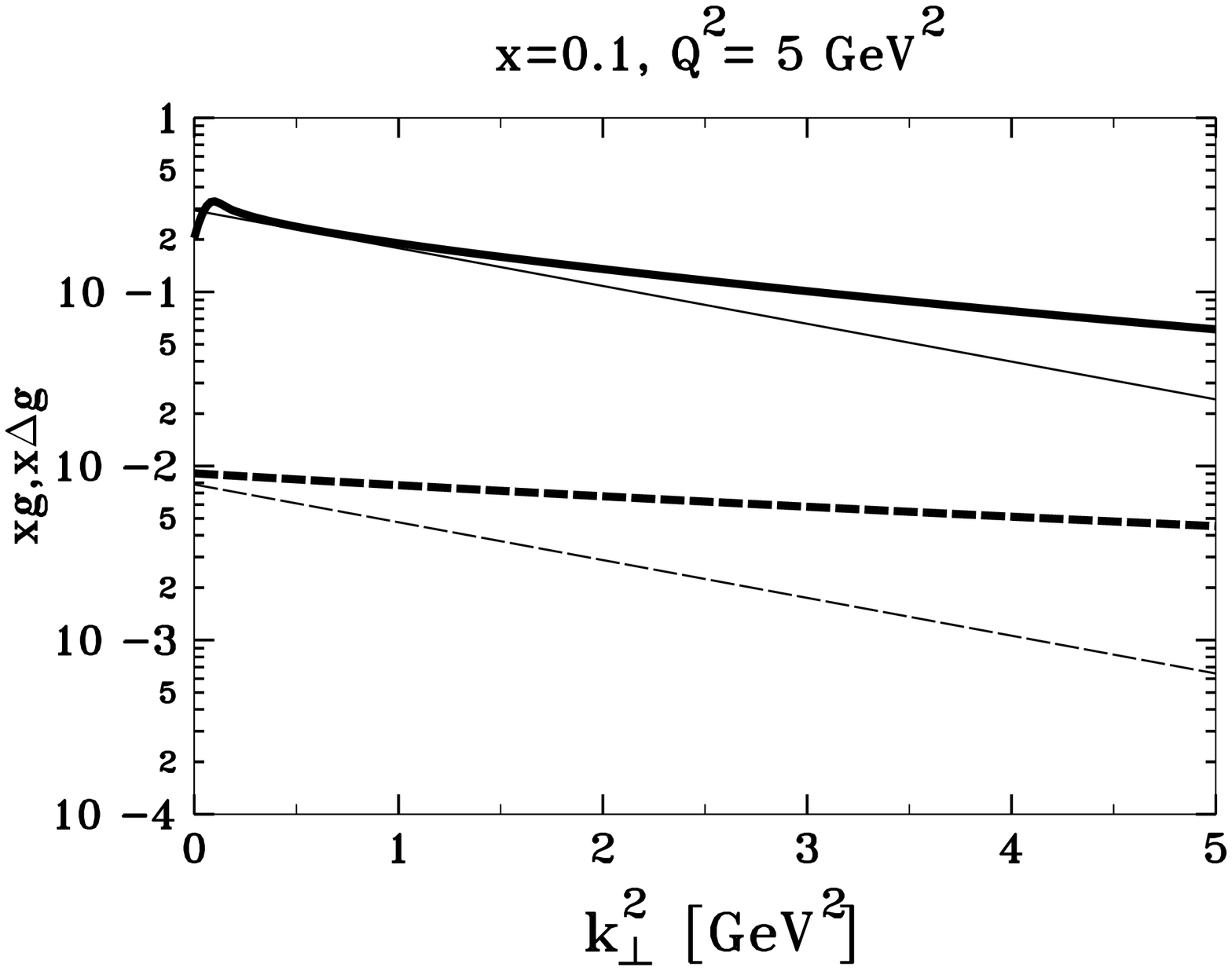,width=10cm}}
\centerline{\psfig{figure=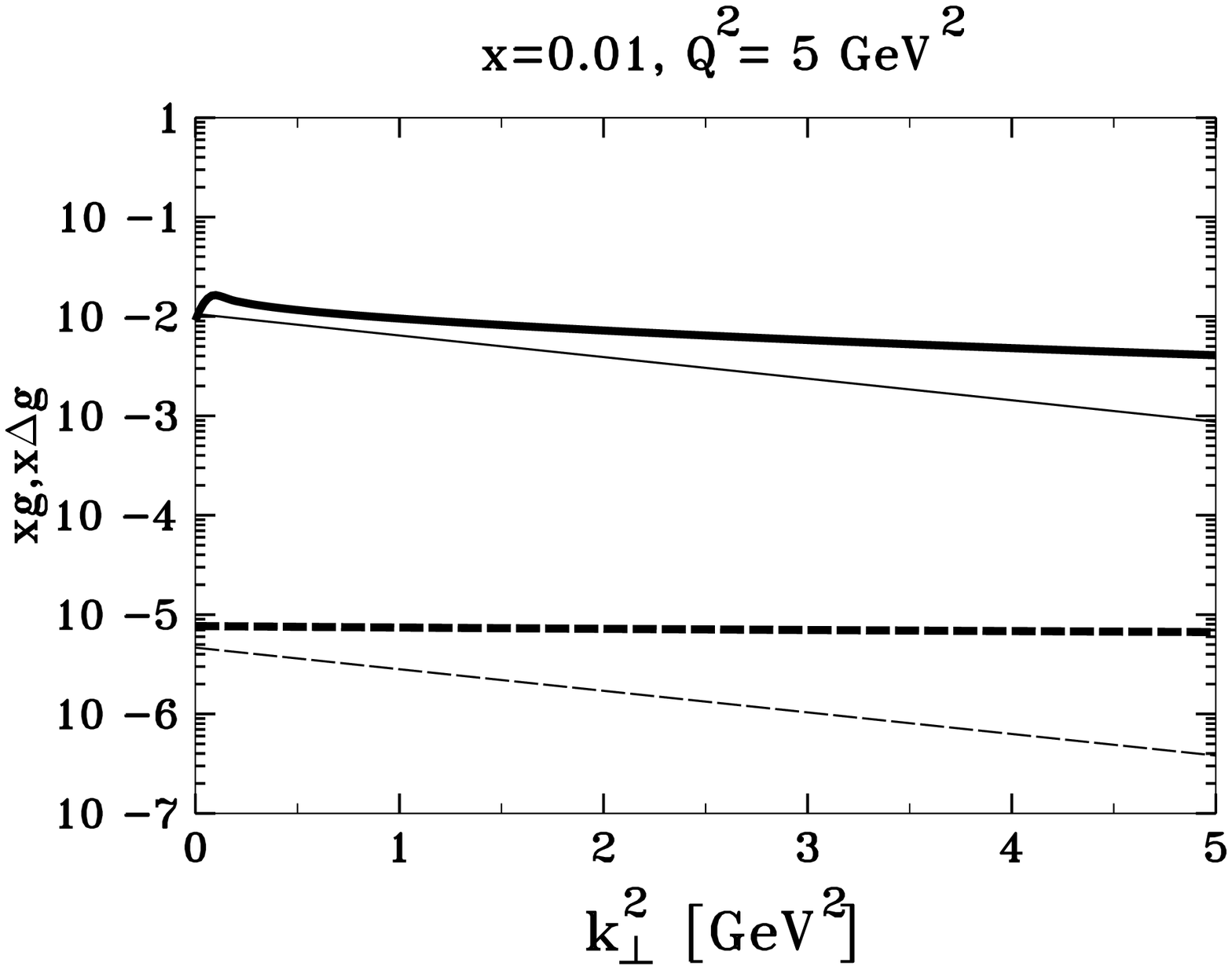,width=10cm}}   
\centerline{\psfig{figure=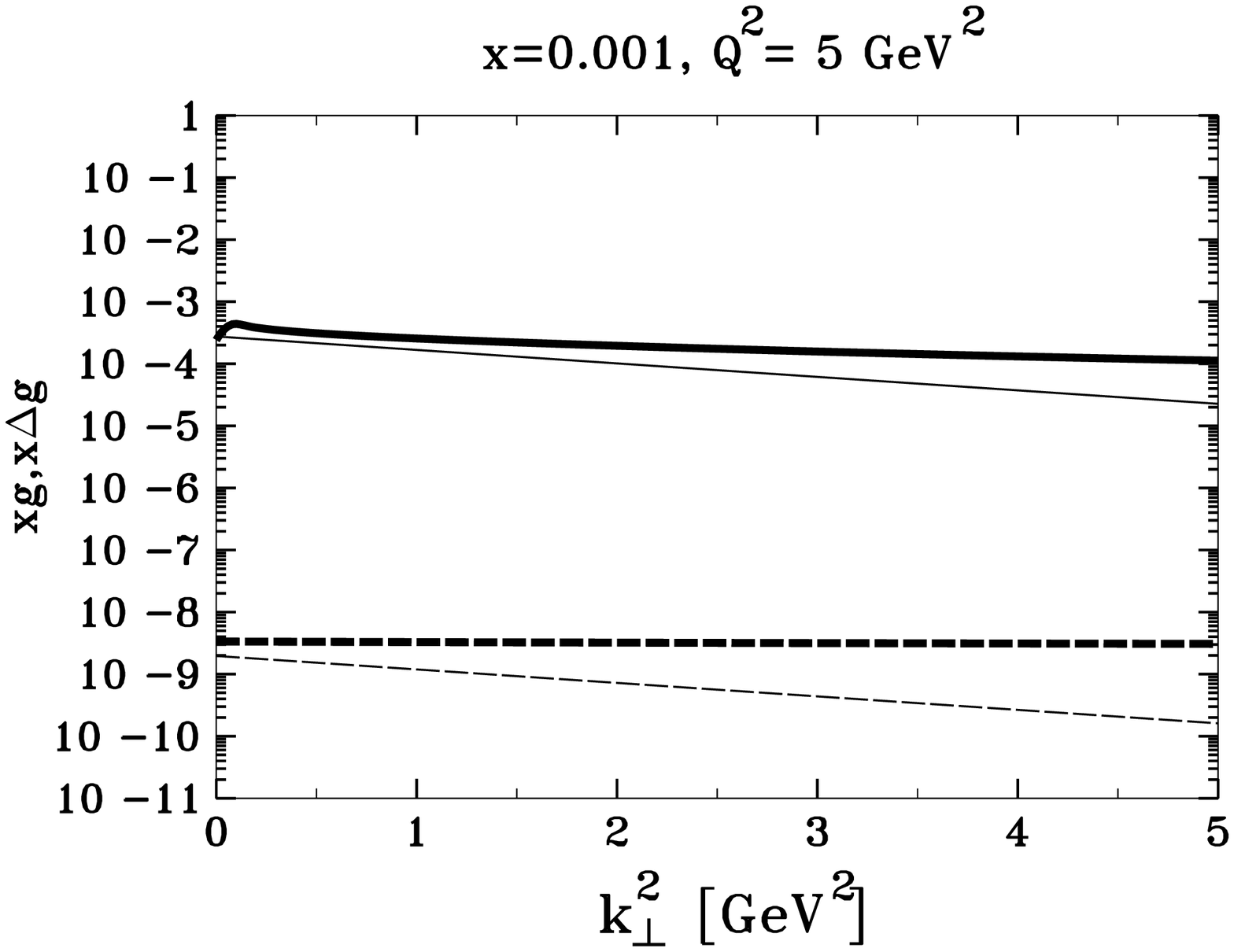,width=10cm}}   
\caption{$k_\perp$ dependence of the unintegrated gluon distribution
$Q^2=5\;{\rm GeV}^2$:\\
bold solid line $xg_1(x,k_\perp^2)$, thin solid line $xg_0(x,k_\perp^2)$,\\
bold dashed line $x\Delta g_1(x,k_\perp^2)$, thin dashed line 
$x \Delta g_0(x,k_\perp^2)$.}
\label{kp2q25}
\end{figure}

As the
CCFM equation is a leading order equation we have to compare it to the 
leading order DGLAP evolution. Consequently, we also have to use for 
the running coupling $\alpha_s$ the leading order formula:
\begin{eqnarray}
\alpha_s(\mu^2) &=& \frac{4\pi}{\beta_0 \ln\left(\mu^2/\Lambda^2\right)},
\qquad \beta_0 = 11-\frac{2}{3}N_f
\nonumber \\
\Lambda_{\overline{\rm MS}}^{(3,4,5,6)}&=& 204, 175, 132, 66.5\;{\rm  MeV}
\nonumber \\
m_{c,b,t} &=& 1.4, 4.5, 175 \;{\rm GeV}\;.
\end{eqnarray}
The values for  $\Lambda_{\overline{\rm MS}}$ and the threshold quark
masses are taken from \cite{Gluck:1998xa}. As a next step we have to
choose the input distribution. With the simplifications done we only
need the x-dependent distributions at the input scale $Q^2_0$. We choose
for the unpolarized gluon distribution GRV(98) LO \cite{Gluck:1998xa}
and in the polarized case GRSV(00) standard scenario \cite{Gluck:2001dy}.
The input scale is in both cases $Q^2_0 = 0.26\;{\rm  GeV}^2$:
\begin{eqnarray}
xg'_{\rm GRV}(x,Q^2_0) &=& 17.47 x^{1.6}(1-x)^{3.8}
\nonumber \\
x\Delta g'_{\rm GRSV}(x,Q^2_0) &=& 1.669 x^{1.79}(1-x)^{0.15} 
(x g'_{\rm GRV}(x,Q^2_0))\;.
\end{eqnarray}
For practical implications it is now also advantageous to transform the
eikonal form factor into a single integral:
\begin{eqnarray}
-\frac{\ln \Delta_{\rm e}^{(g/q)}(q_2^2 ,q_1^2)}{C_{A}} &=& 
\int_{q_1^2}^{q_2^2} \frac{d {q}^2}{{q}^2}
\int_0^{1-Q_0/{q}} \frac{dz}{1-z} \frac{\alpha_s({q}^2(1-z)^2)}{\pi} 
\nonumber \\
&=&
\int_0^{1-\frac{Q_0}{q_1}} \frac{dz}{1-z}
\int_{q_1^2}^{q_2^2}\frac{d q^2}{q^2} 
\frac{\alpha_s\left(q^2(1-z)^2\right)}{\pi} 
+
\int_{1-\frac{Q_0}{q_1}}^{1-\frac{Q_0}{q_2}} \frac{dz}{1-z}
\int^{q_2^2}_{Q_0^2/(1-z)^2}\frac{d q^2}{q^2} 
\frac{\alpha_s\left(q^2(1-z)^2\right)}{\pi} 
\nonumber \\
&=&
\int_0^{1-\frac{Q_0}{q_1}} \frac{dz}{1-z}
\int_{q_1^2(1-z)^2}^{q_2^2(1-z)^2}\frac{d \mu^2}{\mu^2} 
\frac{\alpha_s\left(\mu^2\right)}{\pi} 
+
\int_{1-\frac{Q_0}{q_1}}^{1-\frac{Q_0}{q_2}} \frac{dz}{1-z}
\int^{q_2^2(1-z)^2}_{Q_0^2}\frac{d \mu^2}{\mu^2} 
\frac{\alpha_s\left(\mu^2\right)}{\pi} 
\nonumber \\
&=&
\int_{Q_0^2}^{q_2^2}  \frac{d\mu^2}{\mu}
\frac{\alpha_s(\mu^2)}{\pi} \int_{{\rm max}\left( 0, 1-\frac{\mu}{q_1}\right)}
^{{\rm min}\left( 1-\frac{Q_0}{q_1}, 1-\frac{\mu}{q_2}\right)}
\frac{dz}{1-z} 
+
\int_{Q_0^2}^{Q_0^2\frac{q_2^2}{q_1^2}}  \frac{d\mu^2}{\mu}
\frac{\alpha_s(\mu^2)}{\pi} \int_{ 1-\frac{Q_0}{q_1}}
^{{\rm min}\left( 1-\frac{Q_0}{q_2}, 1-\frac{\mu}{q_2}\right)}
\frac{dz}{1-z} 
\nonumber \\
&=&
\int_{Q^2_0}^{q_1^2} \frac{d\mu^2}{\mu^2} \frac{\alpha_s(\mu^2)}{2\pi}
\ln \left(\frac{\mu^2}{q_1^2}\right)
-
\int_{Q^2_0}^{q_2^2} \frac{d\mu^2}{\mu^2} \frac{\alpha_s(\mu^2)}{2\pi}
\ln \left(\frac{\mu^2}{q_2^2}\right)
\end{eqnarray}
Also the non-eikonal kernel, which is important for the unpolarized case,
can be given an analytical structure:
\begin{eqnarray}
\Delta_{\rm ne} (z,q,k_\perp) &=&
\exp\left[ -\frac{\alpha_s(Q^2(1-z)^2)}{\pi}C_A
\ln(z)\left( \ln z - \ln \frac{k_\perp^2}{q^2}\right)\right]\;. 
\end{eqnarray}
Figs.~\ref{kp2q22} and \ref{kp2q25} show the $k_\perp$ dependence of
the functions $xg_1$ and $x\Delta g_1$ for $Q^2=2\;{\rm GeV}^2$
and $Q^2=5\;{\rm GeV}^2$ for various values of $x$. We have chosen
relatively small values of $Q^2$ because we cannot expect to describe
large $Q^2$ values well by a single dressed gluon emission.
As our input scale $Q_0^2$ is quite low, many dressed gluon emissions
are necessary to get on to a stage where a realistic gluon distribution
is obtained.  Here we only want to see what effect the single 
dressed gluon emission has on the $k_\perp$ distribution.
It is seen that
for small values of $k_\perp$ there is a difference between the polarized
and the unpolarized case due to the fact
that in the polarized case the non-eikonal form factor is absent.
In both cases one sees that the dressed single soft gluon emission
leads to a strong broadening of the $k_\perp$ dependence. To a good
approximation the  $k_\perp$ distribution from the single dressed
gluon emission is again of Gaussian type,
especially in the polarized case where the non-eikonal form factor
is absent.
In general, for large $k_\perp$ the behavior is similar in the unpolarized
case and in the polarized case.
\section{Summary and conclusions}
In this paper we have derived a polarized version for the pure gluon
part of the CCFM equation. Comparing the polarized version
with the unpolarized one the following remarkable features are seen:
\begin{itemize}
\item
We have seen that for $z\to 1$ the
hard splitting kernel in the polarized CCFM equation coincides with 
the $1/(1-z)$ pole in the corresponding polarized DGLAP splitting function.
This is in parallel to the unpolarized case where the $1/(1-z)$ pole
in the hard splitting function P(z) is identical to the $1/(1-z)$ pole
in the corresponding unpolarized DGLAP splitting function.
\item
We have shown that due to  spin correlation
in the polarized case no non-eikonal contributions arise, and that
for the same reason
also the non-eikonal form factor does not enter into the polarized
CCFM equation.
\item
This is consistent with the observation that there is no 1/z pole
in the polarized case for the hard splitting function $\Delta P(z)$ because
the non-eikonal form factor is coupled to that pole. The absence of such
a 1/z pole in $\Delta P(z)$ is furthermore  consistent with the fact
that such a pole does not exist in the corresponding polarized DGLAP
splitting function for the gluons.
\item
Finally, we have shown that considering single dressed gluon emission one
finds  for large $k_\perp$ a similar broadening of the $k_\perp$ 
distribution  as in the unpolarized case. Differences occur only
for small $k_\perp$, where the absence of the non-eikonal form factor becomes
noticeable.
\end{itemize}
Next steps to be taken is a comparison to data. Here the first main 
difficulty will be to solve the polarized CCFM equation which seems
to be possible only in terms of a Monte Carlo simulation of the soft
emission. 
The second problem lies in the fact that the contribution from 
the quarks to the total polarized cross section is in principle of the
same order as the one from the gluons.
Therefore, it will be necessary to extend this formalism to the
polarized contributions of the quarks as well. 
When there are quarks
as initial states, a different type of vertex arises, where the polarization
flow necessary enters into the soft emission of the final quark. 
Non-eikonal contributions become important and the simple reasoning 
of the polarization
flow we discussed for the gluons cannot be applied.
Future work will
therefore have to deal predominantly with those two problems before
a comprehensive comparison to data and a comparison to other
calculations like the one in Ref.~\cite{Bartels:2001zs} can be done.
\newline
\newline
\medskip
I wish to acknowledge useful discussion with B.~Andersson,
G.~Gustafson and L.~L\"onnblad.

\newpage
\end{document}